# Dynamic MDS Matrices for Substantial Cryptographic Strength

Muhammad Yasir Malik and Jong-Seon No

Dept. of EECS, INMS
Seoul National University
`yasir@snu.ac.kr,jsno@snu.ac.kr`

**Abstract.** Ciphers have their strength from the mathematical functions of confusion and diffusion, also known as substitution and permutation. These were the basics of classical cryptography and they are still the basic part of modern ciphers. In block ciphers, diffusion is achieved by the use of maximum distance separable (MDS) matrices. In this paper, we present some methods for constructing dynamic (and random) non-linear MDS matrices.

## 1  Introduction

As defined by Claude Shannon in his paper "Communication Theory of Secrecy Communication", confusion and diffusion are two properties necessitated for the operation of any secure cryptographic algorithm [1]. In the classic ciphers these attributes were achieved by the use of substitution or transposition ciphers. Modern cryptography utilizes the same phenomenon in the form of substitution-permutation network (SPN). Feistel networks employing S-box provide the same effect but at much higher cost in terms of execution units [2][3].

Diffusion in SPN is due to the practice of permutation boxes which dissipate the statistical structure of the plaintext in the long range statistics of the ciphertext. Maximum distance separable (MDS) matrices, which are mainly derived from Reed-Solomon codes, deliver the diffusion properties thus making them one of the vital constituent of modern age ciphers like Advanced Encryption Standard (AES) and Two Fish.

Serge Vaudenay proposed MDS matrices (linear multipermutations) as a cipher design element [5]. Some of the famous ciphers that make use of MDS matrices are AES [6], Two Fish [7], Shark and Square. AES and Two Fish use predefined 4×4 MDS matrices (in field $F_{2^8}$ for computations) for rendering diffusion to the resulting cipher text. Other ciphers utilizing these matrices are Manta, Hierocrypt, and Camellia, and the stream cipher MUGI and the cryptographic hash function WHIRLPOOL.

## 2   Dynamic MDS Matrices

Nearly all the ciphers which incorporate MDS matrices as their mathematical structure for diffusion use predefined MDS matrices. Although in some ciphers the possibility of random selection of MDS matrix with some constraint is provided, but this idea is not being exercised in practical applications.

### 2.1   Dynamic MDS Matrix from Random Constant

A method for generating MDS matrix is proposed in [8]. Already available MDS matrix can be used to reconstruct new nonlinear random matrix.

*Theorem:*

Let $A = [a_{i,j}]_{m \times m}$, $a_{i,j} \in F_q$ be an MDS matrix. Then for an element $e \neq 0 \in F_q$, $eA$ is an MDS matrix.

*Proof:*

We will use the property of non-singularity of sub matrices of $A$. Let $A$ be an MDS matrix for any square sub-matrix $S$ of $A$. From the property of MDS matrices, every sub matrix should be non-singular $|S| \neq 0$. This theorem can be proved by contradiction. We now generate a new matrix $Á$ such that

$$Á = eA, \text{ where } e \neq 0 \in F_q.$$

The new square sub matrix will be

$$Ś = eS, \text{ for all sub matrices } Ś \text{ of } Á.$$

Suppose that $|S'| = 0$, which implies $|eS| = 0$, or $e|S| = 0$. Thus,

$$|S| = 0 \text{ and thus } e \neq 0$$

which contradicts the assumption that S is non-singular. This proves that A is an MDS matrix.

This method can introduce nonlinearity in MDS but at a higher cost for the field $F_{2^8}$. There are a number of multiplications involved (modular multiplication at each step), 256 lookup tables are required. The results can be pre-computed and saved in stack for immediate usage. Before discussing implementations and effectuations of this method, we will discuss and manipulate another method for yielding random and dynamic MDS matrices.

### 2.2   Dynamic MDS Matrix from Optimal Matrix

Junod and Vaudenay proposed efficient MDS matrices by maximizing the number of 1's and minimizing different constants in the matrix [9]. Multiplications by 1 are "free" operations so they can improve the computational efficiency of MDS matrices. MDS matrices need to be bi-regular arrays. A 2×2 array with entries in the field is bi-regular if at least one row and one column of the matrix have two different entries.

Example of an "optimal" bi-regular 4×4 -matrix is

$$\begin{pmatrix} a & 1 & 1 & 1 \\ 1 & 1 & b & a \\ 1 & a & 1 & b \\ 1 & b & a & 1 \end{pmatrix}.$$

The optimal matrix consists of many 1's and a few distinct constants ($a$, $b$ and $c$). It is obvious that multiplying with this method will yield efficient results because of many transparent multiplications with 1's.

For generating new and efficient MDS matrices from the already given matrix, maintaining quite a large number of 1's is required. Let $A$ be the 4×4 optimal matrix as given in the above example. For assuring presence of 1's in the matrix even after multiplication by a random constant $e$, we have to convert any constant in the bi-regular array to 1. It can be done simply by choosing inverse of any constant other than 1 and multiply the given matrix with the calculated inverse. Modular inverse of the constant can be found by

$$a * a^{-1} \equiv 1 \bmod q.$$

The modular multiplicative inverse of modulo $q$ can be found with the extended Euclidean algorithm. The algorithm finds solutions to Bézout's identity

$$ax + by = \gcd(a, b)$$

where $a$, $b$ are given and $x$, $y$, and $gcd(a, b)$ are the integers that the algorithm discovers. This algorithm runs in time $O(\log(m)^2)$, assuming $|a|<m$, and is generally more efficient than exponentiation.

The new form of the generated MDS matrix in $F_{2^8}$ will be

$$a^{-1} * A = \begin{pmatrix} 1 & a^{-1} & a^{-1} & a^{-1} \\ a^{-1} & a^{-1} & a^{-1}b & 1 \\ a^{-1} & 1 & a^{-1} & a^{-1}b \\ a^{-1} & a^{-1}b & 1 & a^{-1} \end{pmatrix}.$$

All occurrences of constant $a$ will convert to 1 which provides the advantage of effective and cheap computation. Moreover the inherited vantages of "free" multiplications also have the positive aspect on the computation efficiency for generation of random MDS matrix. On the other hand, there is a drawback associated with this technique. In [9], it has been showed that the numbers of 1's that can occur in optimal (and partially-optimal) matrices are more than the number of different constants. Considering this fact, it seems that we may not be able to sustain the maximum number of 1's, which is a requirement for efficiency.

## 3   Implementation Considerations and Conditions

This section presents implementation circumstances for constructing dynamic MDS matrices from any given MDS matrix. Two methods are provided in the previous portion of the paper. For practical implementation of these techniques, the most important consideration is the type of matrix which is already present. Type of matrix defines expenditures in computation, time, and power.

### 3.1 Considerations for Different Types of Matrices

For this subsection we consider circulant, non-circulant, optimal, non-optimal, and "worst-case" type matrices for returning dynamic matrices.
AES uses 4×4 circulant matrix [6] whereas Two Fish uses a 4×4 non-circulant matrix [7]. Circulant matrix is a special kind of Toeplitz matrix where each row vector is rotated one element to the right relative to the preceding row vector. In non-circulant matrix, the elements of row vector are repeated but they may not follow any specific arrangements.

We will use the method given in Section 2.1 for generation of dynamic MDS matrix. In this method a randomly selected number $e$ is multiplied with the given matrix. For circular matrices the benefit of repeated constants is present and we can harness this by computing the modular multiplications of the matrix with the random constant, storing it in stack and assigning the memory address to a pointer. For these repeated constants in circulant matrix, we can simply use the pre-computed values instead of doing all the multiplications over and again. The same holds for given non-circulant matrices. Dimension and number of elements in matrices are known and elements are modular multiplied by values of random constant ranging from '01' to any number defined within $F_{2^8}$.

For optimal matrix, many multiplications are saved because of free multiplication of the random constant with 1's. Now we consider the cases of non-optimal matrix and so-called worst-case matrix (with all distinct constants). For optimal matrices, requirement for computation is more than optimal matrices because of lesser 1's and more distinct constants. As these constants are being used in all the row vectors, they play role in saving computations and memory. For matrix with all different constants, the demand for computation and memory is at maximum.

## 4 Performance Analysis

These methods for random MDS matrix generation are implemented on C54xx family series by Texas Instruments. TMS320VC5416 fixed-point digital signal processor is used for this purpose. It is a 16-bit processor with operations at 160 MHz. It incorporates 128K × 16-bit on-chip RAM and 16K × 16-bit on-chip ROM which is configured for program memory [10].

For the method involving multiplicative inverses, before modular multiplication with elements of the existing matrix, the inverses are to be taken. This causes a drawback for efficient usage of this technique. Because of its efficiency in terms of computation, we have implemented method given in Section 2.1. Following table shows the computational requirements for generation of non-linear dynamic MDS matrices. All values are taken for a given 4×4 matrix.

**Table 1.** Performance analysis of different matrices.

| Type of matrix | CPU cycles | Time | Memory requirement |
|---|---|---|---|
| Optimal | 2,077,750 | 13 ms | 762 bytes |
| Circulant | 2,723,924 | 17 ms | 1016 bytes |
| Non-circulant | 2,723,924 | 17 ms | 1016 bytes |
| Non-optimal | 3,394,206 | 21 ms | 1016 bytes |
| Worst case | 4,467,946 | 28 ms | 5588 bytes |

Results show that the optimal matrices are more feasible for formation of new random matrices.

## 5   Applications

Dynamic MDS matrices can be incorporated in block ciphers to ensure security of encryption and decryption. For different communication sessions different random matrices can be utilized. As all the multiplications will be pre-computed so the only random constant is to be chosen. These random constants can be communicated between two parties in a number of ways. If the parties are performing mutual authentication [11], they can share the random constants embedded in their respective digital signatures. Similarly random constants can be treated as message for Standard Hash Algorithm (SHA) function and this hashed value can be exchanged.

After introduction of dynamic MDS matrices, diffusion in the block ciphers will become unpredictable thus enabling strength in the security of the cipher.

## 6   Security Analysis

In this section, immunity of the dynamic MDS matrices against some of the well known cryptographic attacks will be discussed.

MDS matrices are separable and invertible, which causes their computation to be difficult. Their non-linearity makes them resilient against algebraic attacks. Linear cryptanalysis is a known plaintext attack that requires access to large amounts of plaintext and ciphertext pairs encrypted with an unknown key. Differential cryptanalysis is a chosen plaintext attack that seeks to discover a relationship between cipher texts produced by two related plaintexts. It focuses on statistical analysis of two inputs and two outputs of a cryptographic algorithm.

Diffusion caused by the multipermutations (MDS matrices) is non-linear and more difficult to predict because of the usage of random value to generate random MDS matrix. These matrices can be generated dynamically in every round thus causing the prediction of statistical properties and relationships between plain and cipher texts more difficult. Thus linear and differential cryptanalysis can be averted by dynamic generation of MDS matrices.

Side channel attacks (timing attack, power analysis, etc) work by acquiring information from the physical implementation of a cryptosystem. These attacks rely on either emitted electromagnetic radiation or on relationship information (timing and power attacks) [12]. The efficient MDS matrix implementation by Junod and Vaudenay works by maximizing the number of 1's in the matrix. Power used by the operations involving 1's can be used for differential power attacks and thus we need to distribute the Hamming weight to avoid this attack. The methods for making matrices in this paper are more immune towards these attacks because of no dependence on 1's (1$^{st}$ method) and lesser number of 1's (2$^{nd}$ method).

# 7  Conclusion

In this paper we have presented two methods for generating dynamic non-linear MDS matrices. These matrices can be efficiently employed in any of the present block ciphers. These matrices distribute the statistical properties of the plaintext in an unpredictable manner over the cipher text. That's way the cipher becomes immune towards many attacks. These methods can be used with little and no change in the present ciphers. Extra computation is, however, needed for their implementation but there is a benefit of high security associated with this.